# Universities Scale Like Cities

Anthony F. J. van Raan
Center for Science and Technology Studies
Leiden University
Wassenaarseweg 62A
P.O. Box 905
2300 AX Leiden, Netherlands

*Abstract*

*Recent studies of urban scaling show that important socioeconomic city characteristics such as wealth and innovation capacity exhibit a nonlinear, particularly a power law scaling with population size. These nonlinear effects are common to all cities, with similar power law exponents. These findings mean that the larger the city, the more disproportionally they are places of wealth and innovation. Local properties of cities cause a deviation from the expected behavior as predicted by the power law scaling. In this paper we demonstrate that universities show a similar behavior as cities in the distribution of the 'gross university income' in terms of total number of citations over 'size' in terms of total number of publications. Moreover, the power law exponents for university scaling are comparable to those for urban scaling. We find that deviations from the expected behavior can indeed be explained by specific local properties of universities, particularly the field-specific composition of a university, and its quality in terms of field-normalized citation impact. By studying both the set of the 500 largest universities worldwide and a specific subset of these 500 universities -the top-100 European universities- we are also able to distinguish between properties of universities with as well as without selection of one specific local property, the quality of a university in terms of its average field-normalized citation impact. It also reveals an interesting observation concerning the working of a crucial property in networked systems, preferential attachment.*

**Introduction**

Recent work of Bettencourt and West and their colleagues on urban scaling (Bettencourt, Lobo, Helbing, Kühnert and West, 2007; Bettencourt, Lobo, Strumsky and West, 2010) addresses the behavior of cities as complex systems. The most surprising discovery (Bettencourt and West 2010) is that size (in terms of population, number of inhabitants) is the major determinant of most characteristics of a city; history, geography and design have secondary roles. As a result of nonlinear interaction in socioeconomic dynamics, an increase of population size of cities (urban agglomerations ranging from about 100,000 to 10,000,000 inhabitants) leads to a disproportional, power law scaling of important indicators such as economic productivity and innovation capacity, for instance the GMP, the Gross Metropolitan Product. The power law exponents are around 1.15. This means that with every doubling of population, regardless of a city's initial size, the increasing returns to scale (population size) is about 15%. Thus, larger cities produce wealth and new ideas faster. The underlying mechanism is that social interactions become more effective with city size: more economic specialization and division of labor, but at the same time more interconnections and increasing advantages in greater economics of scale, particularly in all kinds of city infrastructure. The power law scaling of socioeconomic city indicators with population size provides the expected behavior. Local characteristics of a city however may cause deviations from the expected behavior.

In this paper we demonstrate that universities show a similar behavior as cities. Universities too are complex systems with their own diverse but interrelated infrastructural facilities, social and economic characteristics. Instead of the Gross Metropolitan Product we use the 'gross university income' in terms of total number of citations; instead of city defined size by population we use university size in terms of total number of publications. We find that deviations from the expected behavior can indeed be explained by local properties of universities.

Earlier relevant work was done by Katz (1999, 2000, 2005) who discussed scaling relationships between number of citations and number of publications across research fields and countries. He concluded that the science system is characterized by a cumulative advantage[1], more particularly a size-dependent 'Matthew effect' (Merton 1968, 1988). This implies a nonlinear increase of impact with increasing size, demonstrated by the finding that the number of citations as a function of number of publications exhibits a power law dependence with an exponent larger than 1. In our previous articles (van Raan 2006a, 2006b, 2008b) we demonstrated a size-dependent cumulative advantage of the correlation between number of citations and number of publications also at the level of research groups and on the level of universities (van Raan 2008a).

**Results and Discussion**

*Data and method*

We used the data of the 500 largest universities in the world (in terms of publications covered by the Web of Science[2]) which we collected for the Leiden Ranking 2011-2012[3]. For our analysis we used the entire data set of the 500 universities worldwide, and a subset of the 100 European top-universities. In this way, we may find differences in the deviations from the expected behavior between the whole set and a subset of universities that are selected on the basis of quality (in terms of citation impact). For a brief but comprehensive and clear explanation of the bibliometric indicators we refer to the methodology section of our Leiden Ranking website and to our recent paper on data collection, indicators, and interpretation of the Leiden Ranking (Waltman et al. 2012). In this study we applied the fractional counting method[4] for English-language papers[5].

---

[1] Cumulative advantage refers to any process in which a specific quantity, mostly a measure of wealth or credit (in our case: citations), is distributed among objects (in our case: publications) according to how much they already have: "the rich get richer". The term cumulative advantage was coined by De Solla Price (1976). In network theory the same process is nowadays referred to as preferential attachment: the more connected a node is, the more likely it is to receive new links.

[2] The Web of Science (WoS) is the comprehensive multidisciplinary publication and citation database produced by Thomson Reuters. Our institute (CWTS) has built an improved and enriched version of the WoS for bibliometric studies.

[3] See http://www.leidenranking.com/. The Leiden Ranking 2011-2012 is based on publications in the period 2005-2009, and citations to these publications in the period 2005-2010. There will be an update of the Leiden Ranking in April 2013.

[4] The Leiden Ranking supports two counting methods: full counting and fractional counting. The full counting method gives equal weight to all publications of a university. The fractional counting method gives less weight to collaborative publications than to non-collaborative ones. For instance, if the address list of a publication contains five addresses and two of these addresses belong to a particular university, then the publication has a weight of 0.4 in the calculation of the bibliometric indicators for this university. The fractional counting method leads to a more proper normalization of indicators and to fairer comparisons between universities. Fractional counting is therefore regarded as the preferred counting method in the Leiden Ranking.

[5] We distinguish between the options 'all' WoS covered publications and 'English publications only'. Comparing the impact of non-English language publications with the impact of publications written in English may not be considered fair. Non-English language publications can be read only by a small part of the scientific community, and therefore these publications cannot be expected to receive similar numbers of citations as publications written in English. Particularly for German and French univer-

First we analyzed the entire set of 500 universities worldwide. We took the following approach. We determined the correlation between the absolute number of citations **C**, considered as the 'gross university income', and the absolute number of publications **P**, considered as 'size' of a university. This yields the average scaling behavior of **C** for a university with size **P**. Next, we determined the deviations of from this average for each individual university by calculating the residuals of the scaling distribution in a similar way as Bettencourt *et al* (2010) did for urban scaling. Finally, we investigated with which bibliometric indicators the residuals correlate. The same procedure is followed with a slightly different scaling in which the number of citations is replaced by the field-normalized number of citations.

### *500 largest universities worldwide*

The first analysis is carried out with the entire set of the 500 largest universities worldwide. We find a power law relation between citations and publications of a university, i.e., a nonlinear relation given by

$$\boldsymbol{C(P)} = \boldsymbol{a_1} \cdot \boldsymbol{P}^{\boldsymbol{\beta_1}} \qquad (1)$$

with exponent $\boldsymbol{\beta_1}$ = 1.25, see Fig. 1. The coefficient $\boldsymbol{a_1}$ with value 0.77 is a normalization constant. As discussed above also on lower aggregation levels, particularly research groups, we find similar scaling properties (van Raan 2006a, 2006b, 2008b). We found that in the case of full counting and all papers (so not only the English-language papers) the exponent is 1.30.

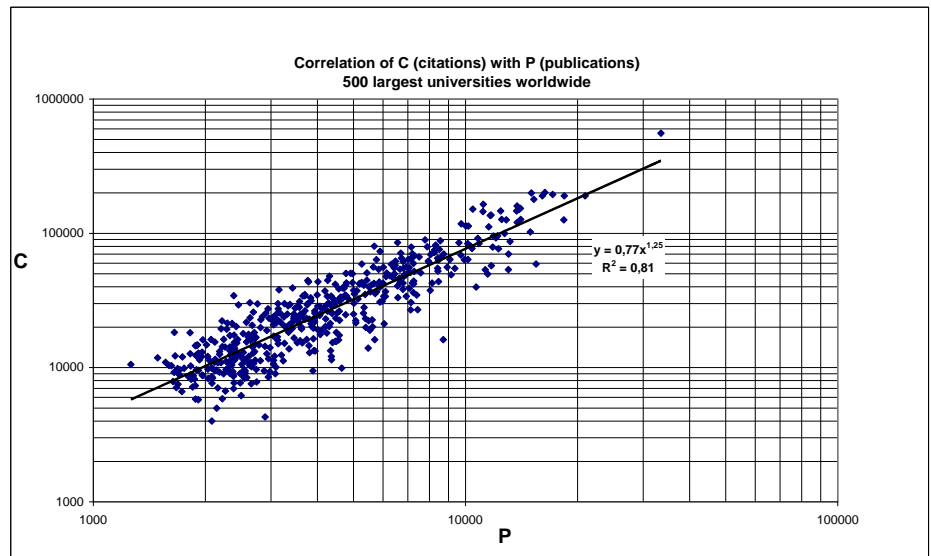

*Fig. 1: Correlation of **C** (citations without self-citations, 'income') with **P** ('size') for the 500 largest universities worldwide.*

Denoting the observed value of the number of citations for each specific university with $\boldsymbol{C}_i$ we calculated the residuals of the scaling distribution of each university in a similar way as Bettencourt *et al* (2010) did for urban scaling:

$$\boldsymbol{\xi_{1i}} = \ln [\boldsymbol{C}_i / \boldsymbol{C(P)}] = \ln [\boldsymbol{C}_i / \boldsymbol{a_1} \cdot \boldsymbol{P}^{\boldsymbol{\beta_1}}] \qquad (2)$$

---

sities this has a strong negative effect on their citation impact and with that on their ranking positions. Therefore the Leiden Ranking 2011-2012 offers the possibility of excluding non-English language publications from the calculation of the bibliometric indicators next to using all publications. For an extensive study of this language effect on university rankings we refer to Van Raan, van Leeuwen and Visser (2011a,b).

Because we will also calculate the residuals of a slightly different scaling, we call the above residuals the C-residuals. The residuals are a measure of the deviation of the real (observed) from the predicted number of citations for each university. In Fig. 2 we show the ranking of the C-residuals for the 500 universities.

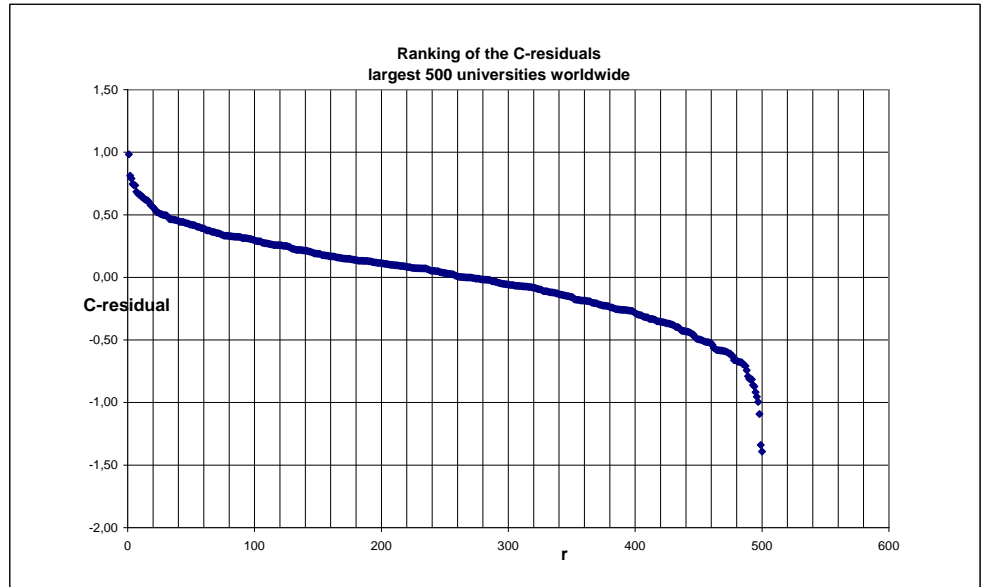

*Fig. 2: Ranking of the C-residuals of the 500 universities*

The central question is what causes the deviations. Particularly, which universities have positive while other have negative residuals. Therefore our next step is to find out with which bibliometric indicators the residuals correlate. We first analyzed the relation with the average number of citations per paper[6] (**MCS**) and with the average *field-normalized* number of citations[7] (**MNCS**) of a university, see Figs. 3 and 4. Notice that the correlation in both cases is not determined by a power law, but by a logarithmic relation. This is what we can expect given the definition of the residuals (see eq. 2). We find:

$$\xi_1 = b_1 \cdot \ln(MCS) - c_1 \quad \text{with } b_1 = 0.86 \text{ and } c_1 = 1.56 \quad (3)$$

and

$$\xi_1 = d_1 \cdot \ln(MNCS) - e_1 \quad \text{with } d_1 = 1.07 \text{ and } e_1 = 0.04 \quad (4)$$

---

[6] *Mean citation score (MCS)*, the average number of citations of the publications of a university.

[7] *Mean normalized citation score (MNCS)*, the average number of citations of the publications of a university, normalized for field differences, publication year, and document type. An **MNCS** value of two for instance means that the publications of a university have been cited twice above world average.

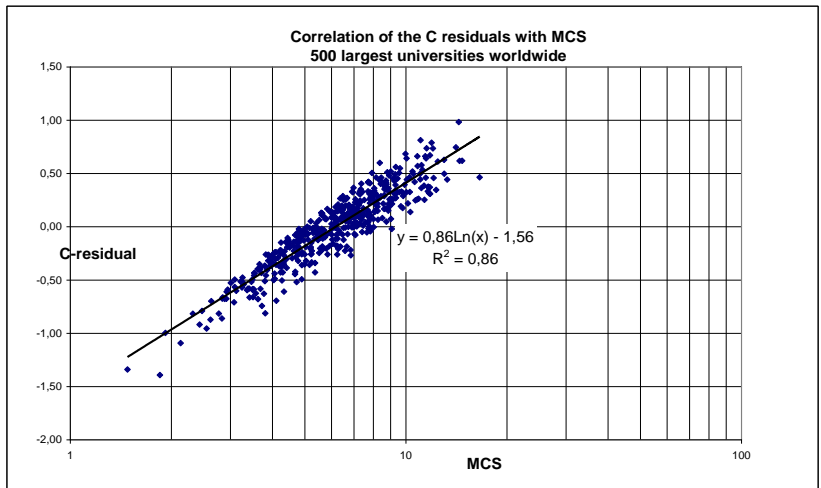

*Fig. 3: Correlation of the C-residuals with the average number of citations per paper for each university (**MCS**) for the 500 largest universities worldwide.*

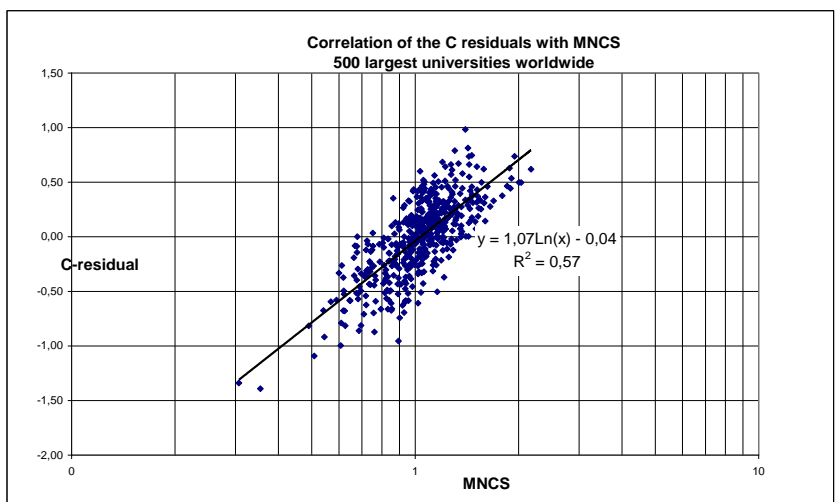

*Fig.4: Correlation of the C-residuals with the average number of field-normalized citations for each university (**MNCS**) for the 500 largest universities worldwide.*

The **MCS** indicator is simply citations per paper: it does not take into account the often large differences in citation density of the many scientific fields. As we showed in earlier work (van Raan 2008a) the average number of citations correlates very strongly with the average field-citation density of a university. In other words, if a university is characterized by strong research activities in fields with the high citation density, and these are predominantly the (bio)medical and the basic natural science fields, then its **MCS** will be relatively high. In contrast, technical universities focus on engineering and applied research, in these fields the citation density is considerably lower as compared to the medical fields, and thus the average number of citation paper for these universities will be lower. The same is true for universities focusing on economics and social sciences.

However, the strong relation of the C-residuals with **MCS** is quite trivial: the C-residuals are a function of the real number of citations and the real number of publications of a university, and the same is true for **MCS**. If the power law exponent would be exactly 1, the correlation between the C-residuals and **MCS** would be perfect. Thus, we need a more convincing prove that the C-residuals are related with the field composition of a university. Therefore we analyzed the relation between the C-residuals and **MCS**/**MNCS**, see Fig. 5. Because **MCS** is the average number of citations per paper and **MNCS** the average field-normalized number of citations of a university, the ratio of both indicators is measure of the average field-citation density of a university, and hence a measure of its field composition.

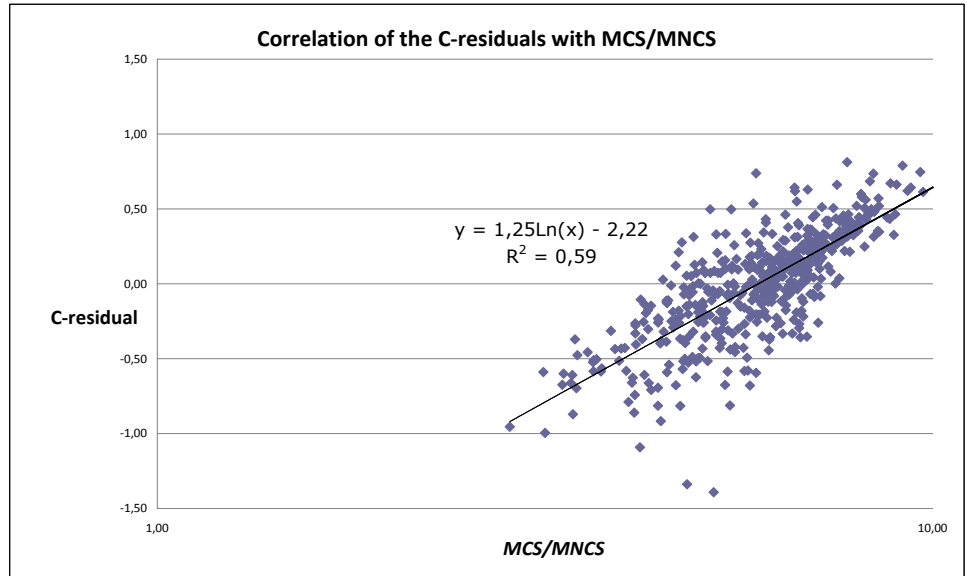

*Fig.5: Correlation of the C-residuals with the average field citation-density for each university (**MCS**/**MNCS**) for the 500 largest universities worldwide.*

Fig. 5 shows the dependence of the C-residuals on the average field citation-density and thus the field-composition of a university, which is a typical 'local characteristic'. In line with our earlier discussion we expect that universities with a large medical school will generally have positive C-residuals, whereas universities with a strong focus on engineering or on social sciences will generally have negative C-residuals. Inspection of the lists of universities with positive and negative C-residuals confirms this assumption.

The **MNCS** indicator corrects for field density, it is the field-normalized impact of a university. As we see in Fig. 4, the C-residuals also correlate reasonably with **MNCS**. This implies that the residuals also reflect to some extent the research quality of a university (measured in terms of our field-normalized citation impact indicator).

In order to approach the problem of field-normalization in a slightly different way, we analyzed the correlation of the field-normalized absolute number of citations ($C_n$ = **P*MNCS**) with the absolute number of publications (**P**). That **P*MNCS** is the field-normalized absolute number of citations can be seen in the following way: **MNCS** indicates the average field-normalized citation per publication and thus by multiplying with the total number of publications, the product is the field-normalized absolute number of citations.

The results are shown in Fig.6. We find a power law relation between field-normalized citations and publications of a university, i.e., a nonlinear relation given by

$$C_n(P) = a_2 \cdot P^{\beta_2} \qquad (5)$$

with exponent $\beta_2$ = 1.17 and coefficient $a_2$ = 0.26 (for full counting and all papers the exponent is 1.19). We notice that the power law exponent in this case is lower than in the case of the non-normalized number of citations (eq. 1).

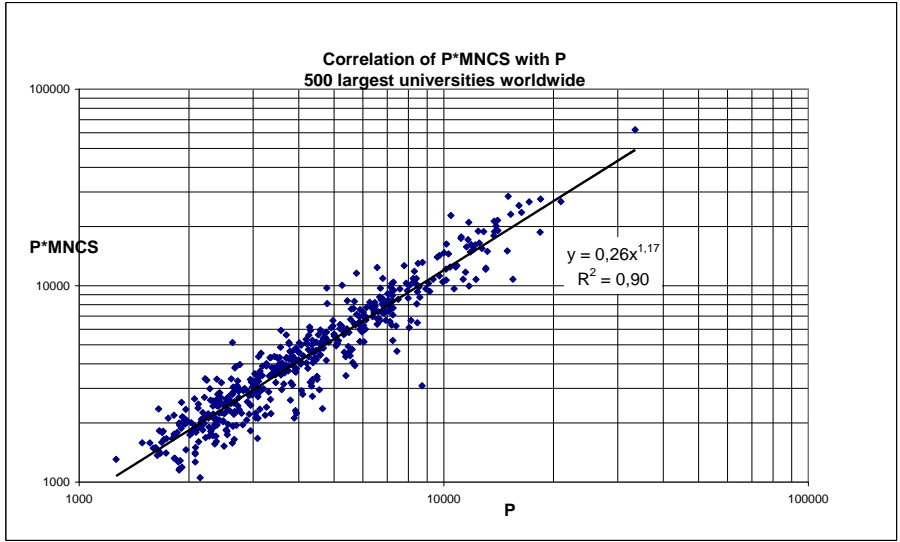

*Fig. 6: Correlation of $C_n$ (= $P*MNCS$, the field-normalized number of citations) with $P$ for the 500 largest universities worldwide.*

Also in this approach we calculated the residuals for each university:

$$\xi_{2i} = \ln[C_{ni}/C_n(P)] = \ln[C_{ni}/a_2 \cdot P^{\beta_2}] \tag{6}$$

We call these residuals the B-residuals. In Fig. 7 we show the ranking of the B-residuals for the 500 universities. We see that the B-residuals cover a somewhat smaller range as compared to the C-residuals.

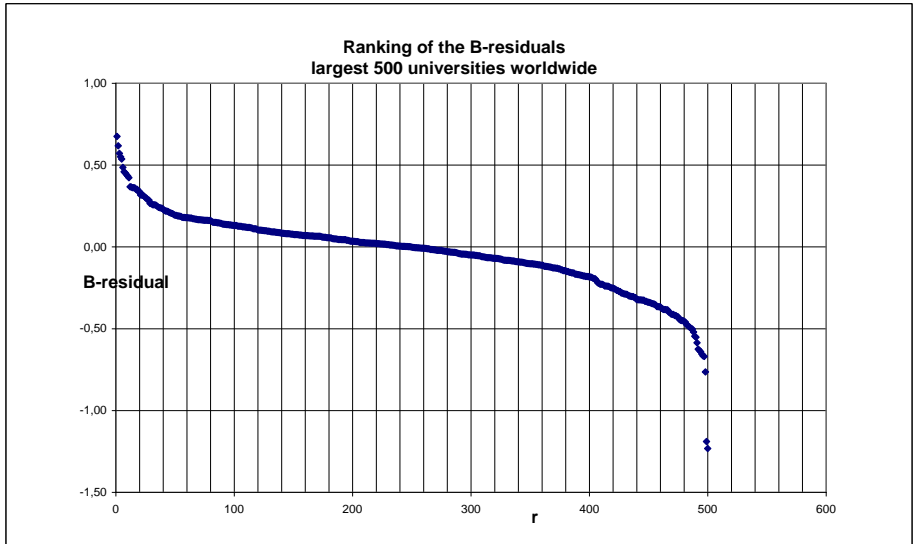

*Fig. 7: Ranking of the B-residuals of the 500 universities.*

The correlation of B-residuals with the average number of citations per paper (*MCS*) of a university and the average field-normalized number of citations (*MNCS*) of a university is shown in Figs. 8 and 9. Again we find a logarithmic relation:

$$\xi_2 = b_2 \cdot \ln(MCS) - c_2 \quad \text{with } b_2=0.45 \text{ and } c_2=0.85 \tag{7}$$

and

$$\xi_2 = d_2.\ln(MNCS) - e_2 \quad \text{with } d_2=0.85 \text{ and } e_2=0.06 \tag{8}$$

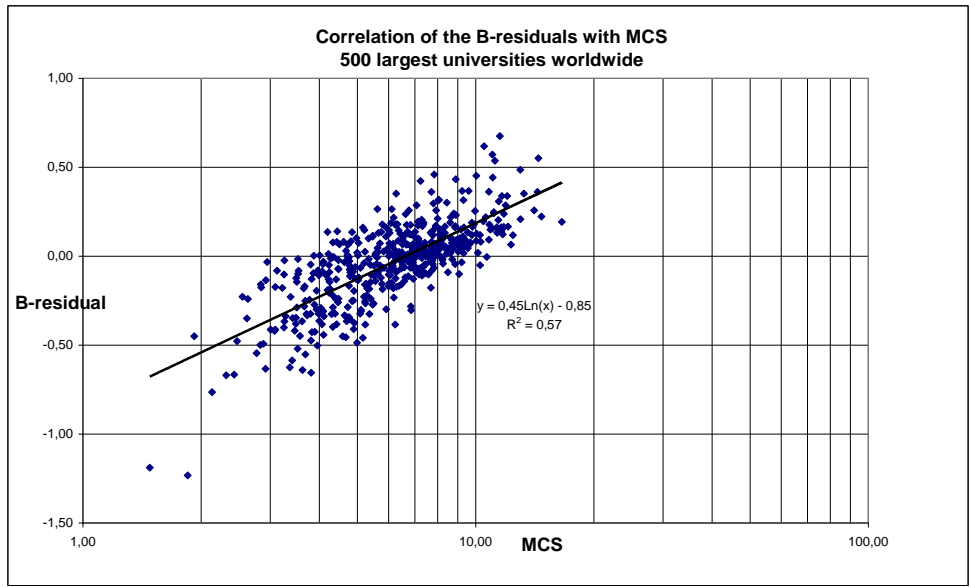

*Fig. 8: Correlation of the B-residuals with the average number of citations per paper for each university (**MCS**) for the 500 universities.*

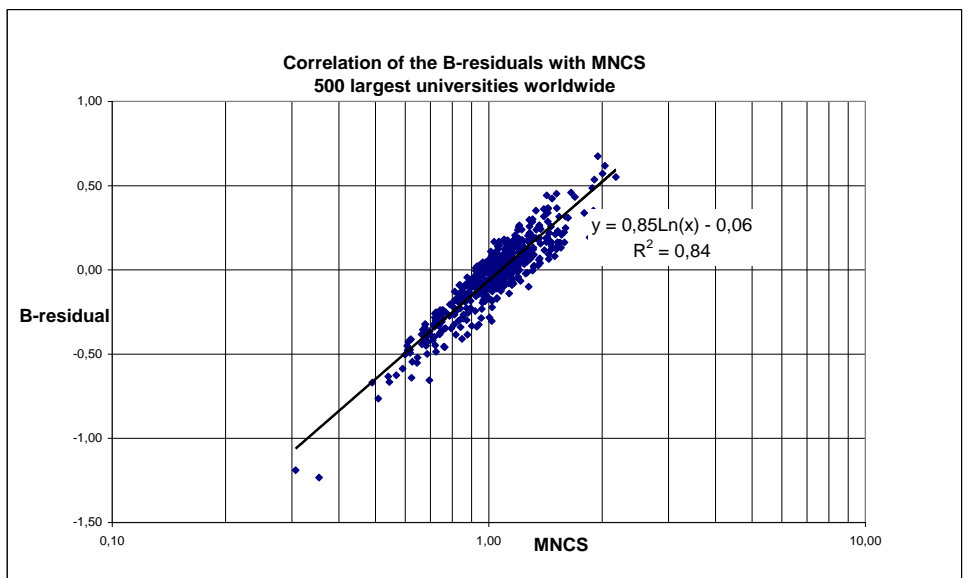

*Fig.9: Correlation of the B-residuals with the average number of field-normalized citations for each university (**MNCS**) for the 500 universities.*

Because the B-residuals are field-normalized measures, we can expect that they will be sensitive for the non-normalized **MCS**. The B-residuals correlate stronger with **MNCS** than in the case of the non-normalized citation scaling. The reason is the same as for the strong correlation of the C-residuals with **MCS**: the B-residuals are a function of the real number of field-normalized citations and the real number of publications of a university, and the same is true for **MNCS**.

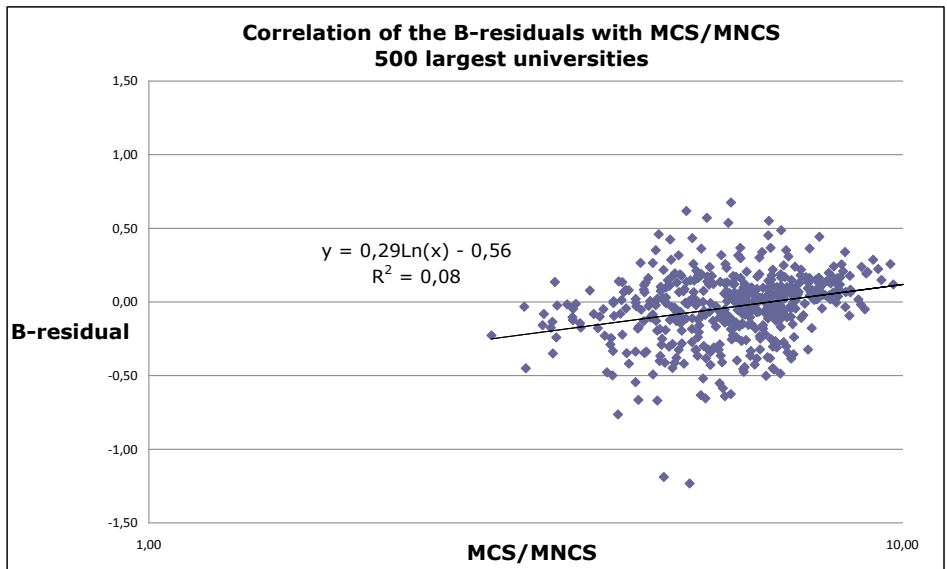

*Fig.10: Correlation of the B-residuals with the average field citation-density for each university (**MCS**/**MNCS**) for the 500 largest universities worldwide.*

There is no significant relation between the B-residuals and the field citation-density indicator **MCS**/**MNCS**, see Fig. 10. This is to be expected because the B-residuals are field-normalized.

Finally, we analyzed the correlation between both types of residuals. As shown in Fig. 11, this correlation is quite significant.

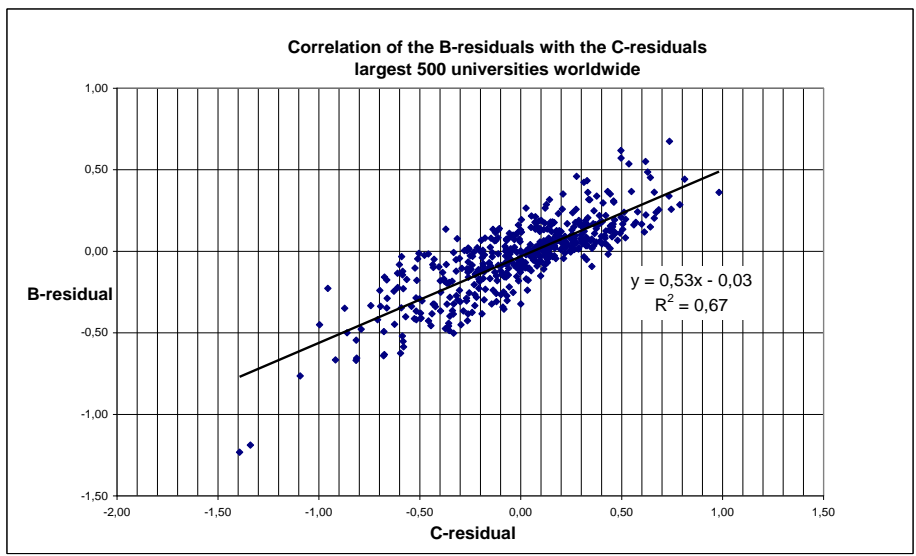

*Fig.11: Correlation of the B-residuals with the C-residuals.*

As remarked earlier, we also performed the entire above analysis for the 500 universities with full counting and all WoS publications. We found that there are no significant differences in results for these two counting modalities.

### *Top-100 European universities*

The selection of the 100 European top-universities was carried out as follows. From the 500 universities worldwide we first identified all European universities (in total 216 of the 500), ranked them by the **PPtop10%** indicator[8], and took the top-100.

Again we find a power law relation between citations and publications of a university, but with a considerably lower exponent as compared to the 500 universities:

$$C(P) = a_3 \cdot P^{\beta_3} \tag{9}$$

with exponent $\beta_3$ = 1.14, and the coefficient $a_3$ = 2.50 , see Fig. 12.

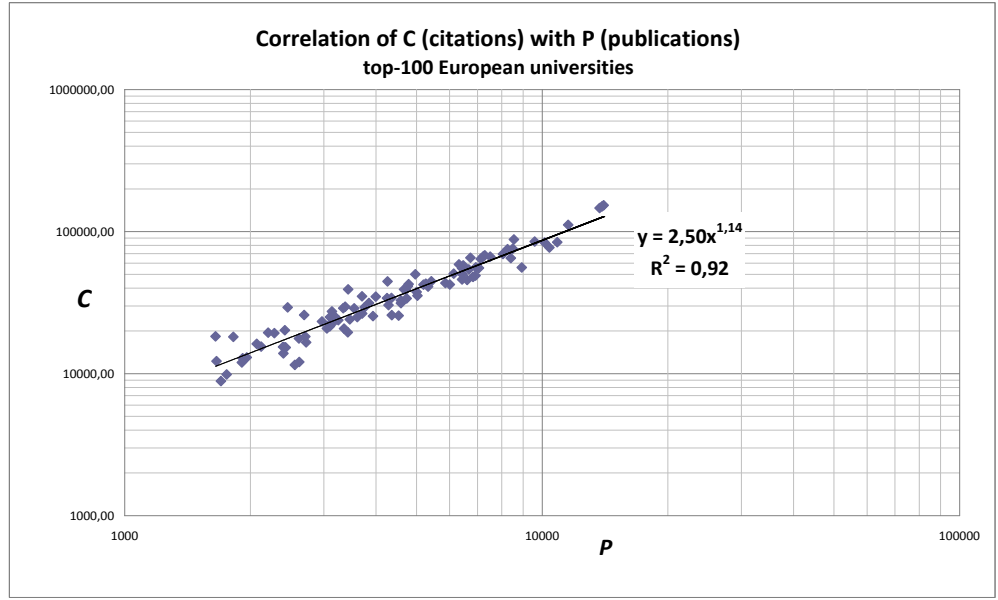

*Fig. 12: Correlation of **C** (citations without self-citations, 'income') with **P** ('size') for the top-100 European universities.*

Our first observation is that for top-universities the non-linear, cumulative advantage by size is considerably less as compared to the set of 500 universities without any quality selection. This is an interesting observation in the context of network theory: if a networked system (here: a university) is characterized by a significantly high number of attractive nodes (here: relatively highly cited publications), then there is 'less to be preferred' (because most of the nodes 'are attractive') and thus the mechanism of preferential attachment will be less strong.

We calculated the C-residuals (eq.10) and in Fig. 13 we show the ranking of these C-residuals for the top-100 European universities.

$$\xi_{3i} = \ln [C_i / C(P)] = \ln [C_i / a_3 \cdot P^{\beta_3}] \tag{10}$$

---

[8] **PPtop10%** *indicator*: the proportion of the publications of a university that, compared with other similar publications, belong to the top 10% most frequently cited of the field(s) to which the publication belongs. Publications are considered similar if they were published in the same field and the same publication year and if they have the same document type. In this way **PPtop10%** is a field-normalized indicator. We further refer to our Leiden ranking site http://www.leidenranking.com/.

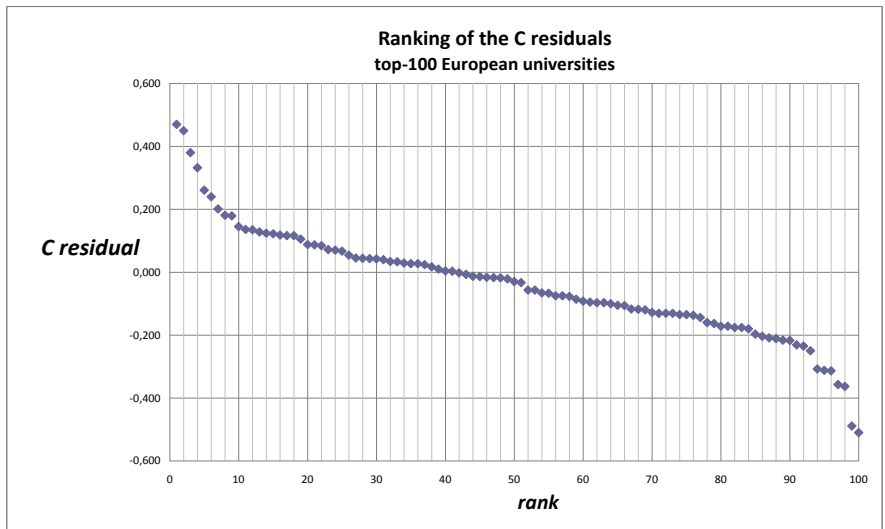

*Fig.13: Ranking of the C-residuals of the top-100 European universities.*

We notice that the range of the C-residuals in the case of the top-100 European universities (between +0.5 and -0.5) is considerably smaller than in the case of the 500 universities (between +1.0 and -1.5). This difference can be explained by the fact that the top-100 European universities are a quality- (citation impact) based selection within the entire set of 500 universities.

The correlation of C-residuals with the average number of citations per paper (**MCS**) of a university and the average field-normalized number of citations (**MNCS**) of a university is shown in Figs. 14 and 15.

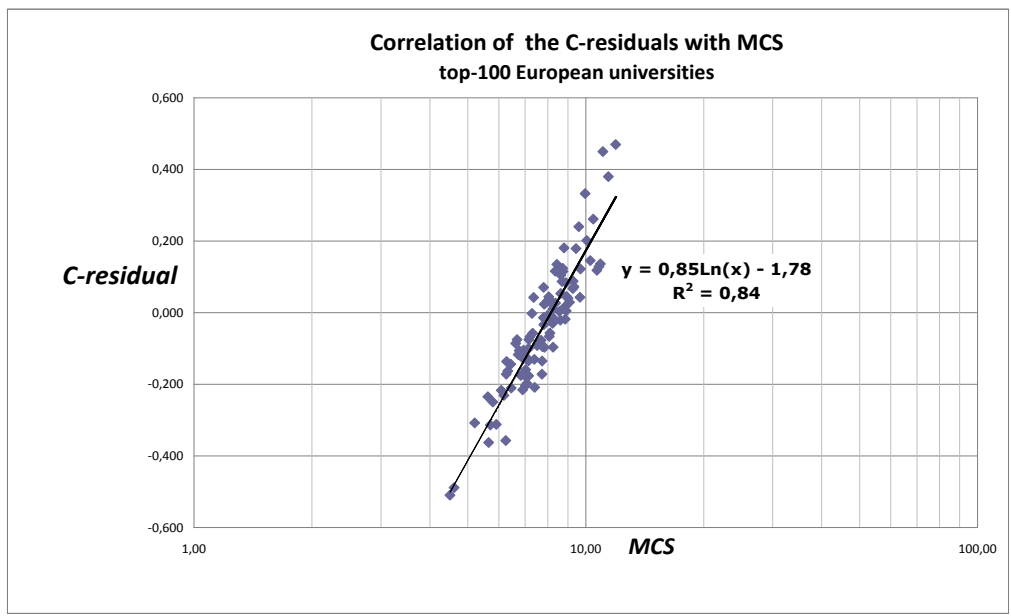

*Fig. 14: Correlation of the C-residuals with the average number of citations per paper for each university (**MCS**) for the top-100 European universities.*

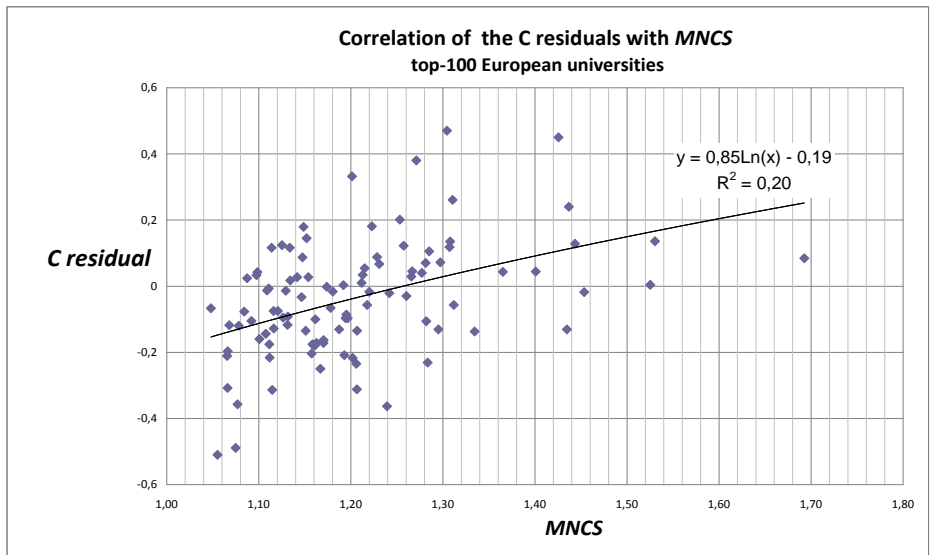

*Fig. 15: Correlation of the C-residuals with the average number of citations per paper for each university (**MNCS**) for the top-100 European universities.*

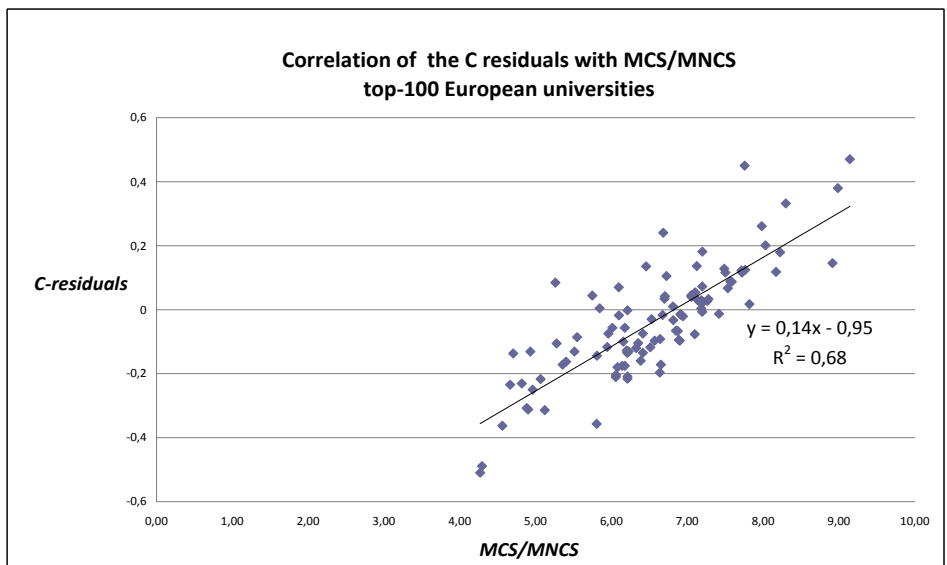

*Fig.16: Correlation of the C-residuals with the average field citation-density for each university (**MCS**/**MNCS**) for the top-100 European universities.*

The correlation of the C-residuals of the top-100 European universities with **MCS** is significantly strong. As discussed earlier, this is to be expected. This correlation however does not significantly follow a logarithmic and can be fit with a linear relation as well.

In complete contrast to our findings for the 500 universities we see in Fig. 15 that the correlation of the C-residuals with **MNCS** is hardly significant. This absence of a significant correlation with the **MNCS** indicator is also what we can expect: the top-universities are all selected on the basis of high impact research and so this indicator will in this specific case hardly discriminate between these universities.

As in the case of the 500 universities, we analyzed the relation between the C-residuals and the average field citation density of a university **MCS**/**MNCS**, see Fig.16. We see a quite significant relation which again implies that the C-residuals are related with the average field citation-density and thus with the field composi-tion of a university. Indeed, also for the top-100 European universities we find by inspecting the C-residuals of the individual universities, that universities with a large medical school generally have positive C-residuals. Universities with a focus on engineering or social sciences generally have negative C-residuals.

In order to find a relation between the C-residuals and other bibliometric indicators we investigated their relation with (1) the percentage of publications in collaboration, (2) the percentage of publications in international collaboration, and (3) the percentage of publications in long distance collaboration[9]. We find that there is no significant relation between the C-residuals and any of these scientific collaboration indicators.

Like in the case of the 500 universities, we also analyzed the correlation of the field-normalized absolute number of citations ($C_n = P*MNCS$) with the absolute number of publications ($P$). The results are shown in Fig.17. We find a power law relation given by

$$C_n(P) = a_4 \cdot P^{\beta_4} \qquad (11)$$

with exponent $\beta_4$ = 1.04 and coefficient $a_4$ = 0.85. This exponent is considerably lower than in the case of the scaling of citations with publications and even more, this power law exponent is almost 1 so that in this case we cannot speak of 'advantage' of size. This is remarkable contrast with the 500 universities where the exponent of the correlation between $P*MNCS$ with $P$ is 1.17 and thus a non-linear, cumulative advantage by size is present. Evidently, the size-dependent cumulative advantage mechanism is considerably weaker in the case of high impact universities.

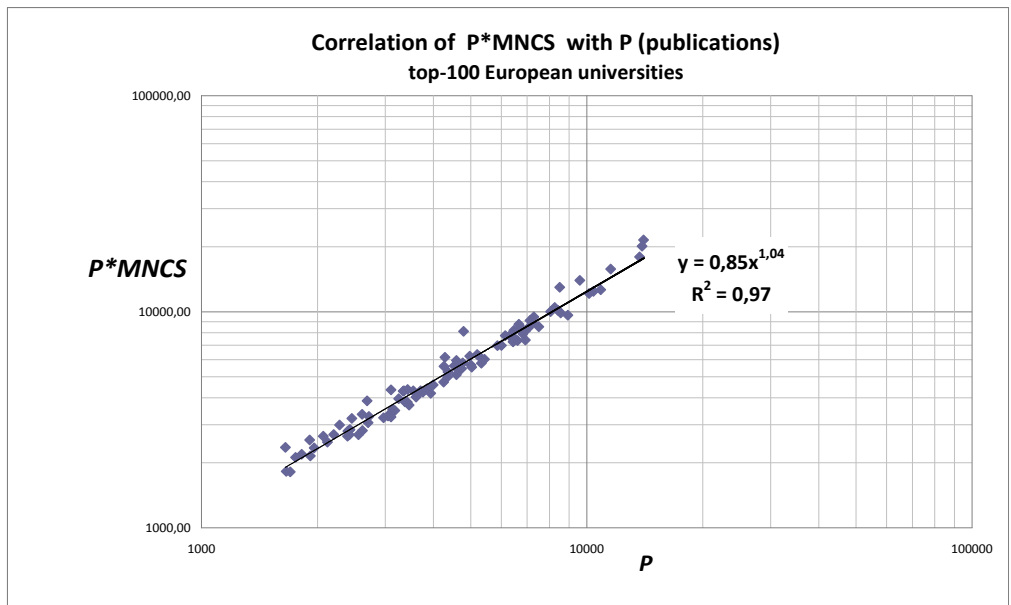

Fig. 17: Correlation of $C_n$ (= $P*MNCS$, the field-normalized number of citations) with $P$ for the top-100 European universities.

In order to avoid the paper becoming lengthy by presenting all further similar details of the B-residuals as in the above analyses[10], we restrict ourselves to the following observations. The B-residuals of the top-100 European universities do not show a correlation with **MCS** (in the case of the 500 universities there is a reasonable correlation); top-universities generally score high on **MCS** and thus this indicator discriminates only to a small extent between the top-universities. The B-residuals correlate stronger with **MNCS** than in the case of the 500 universities. As discussed earlier, the B-residuals are a function of the real number of field-normalized citations and the real number of publications of a university, and the same is true for **MNCS**. Furthermore, the power exponent of the correlation of **P\*MNCS** with **P** is 1.04, so very close to one, which will make the correlations of the B-residuals and **MNCS** almost perfect. As opposed to the 500 universities, the B-residuals of the top-100 European universities show only a very weak to hardly any correlation with the C-residuals.

---

[9] These collaboration indicators are also part of the Leiden Ranking 2011-2012, see http://www.leidenranking.com/methodology.aspx.
[10] These details can be obtained from the author.

**Conclusions**

In this paper we demonstrate that universities show a similar behavior as cities in the distribution of the 'gross university income' in terms of total number of citations over 'size' in terms of total number of publications. Moreover, the power law exponents for university scaling are comparable to those for urban scaling. Also similar to urban scaling, the deviations from the expected behavior can be explained by specific local properties of universities, particularly the field-specific composition of a university, and its quality in terms of field-normalized citation impact.

By studying both the set of the 500 largest universities worldwide and a specific subset of these 500 universities namely the top-100 European universities we are able to distinguish between properties of universities with as well as without selection of one specific local property, the quality of a university in terms of its average field-normalized citation impact. We find a reasonably significant relation of the C-residuals with the average *field-normalized* number of citations of a university (**MNCS**) in the case of the 500 largest universities worldwide, and hardly a significant relation in the case of the top-100 European universities. This implies that the C-residuals are related with the quality (impact) of a university in the case that the set of universities under study is not selected on the basis of quality. If this is the case, such as the top-100 universities, it is obvious that the dependence of the residuals on quality largely disappears.

We also find a reasonably significant relation of the C-residuals with the average field-citation density of a university (**MCS**/**MNCS**), and hence a measure of its field composition, which implies that the C-residuals are related with the average field citation-density and thus with the field composition of a university. We observe this relation in both the set of the 500 largest universities worldwide and in the subset of the top-100 European universities. The analyses of the B-residuals clearly confirm confirms the role of quality as (one of ) the causes for the deviation of universities from the average (expected) scaling behavior.

In summary, our results show that the C-residuals as a measure of deviation from the average scaling behavior relate to two crucial 'local properties' of a university. First, the *citation density of the fields* covered by a university, and thus the field specific composition of a university. For instance, the negative residuals (fewer citations than expected) are characteristic for universities of technology and universities with large engineering departments. And the opposite, the positive residuals are typical for universities with large medical schools. Second, the residuals also relate to the *quality* of a university measured in terms of overall field-normalized citation impact.

Finally, our comparison of sets of universities with and without selection on the basis of citation impact reveals an interesting observation concerning the working of a crucial property in networked systems, preferential attachment. If a networked system (here: a university) is characterized by a significantly high number of attractive nodes (here: relatively highly cited publications), then there is 'less to be preferred' (because most of the nodes 'are attractive') and thus the mechanism of preferential attachment will be less strong.

*Acknowledgements*

I thank Ludo Waltman for his work on the Leiden Ranking data and indicators and for stimulating discussions.